\documentclass[%
 aip,
 amsmath,amssymb,
 reprint,%
]{revtex4-1}

\usepackage{graphicx}
\usepackage{dcolumn}
\usepackage{bm}

\usepackage[utf8]{inputenc}
\usepackage[T1]{fontenc}
\usepackage{mathptmx}
\usepackage{etoolbox}
\usepackage	[colorlinks,	
linkcolor=blue,
anchorcolor=blue,
citecolor=blue]{hyperref}	
\makeatletter
\def\@email#1#2{%
 \endgroup
 \patchcmd{\titleblock@produce}
  {\frontmatter@RRAPformat}
  {\frontmatter@RRAPformat{\produce@RRAP{*#1\href{mailto:#2}{#2}}}\frontmatter@RRAPformat}
  {}{}
}%
\makeatother

\draft 

\begin{document}

\preprint{AIP/123-QED}

\title{Practical ultra-low frequency noise laser system for quantum sensors}
\author{Shiyu Xue}%
\author{Mingyong Jing} 
\email{jmy@sxu.edu.cn}
\thanks{These authors contributed equally to this work}
\altaffiliation[Also at ]{State Key Laboratory of Precision Measurement Technology and Instruments, Department of Precision Instrument, Tsinghua University, Haidian, Beijing 100084, China}

\author{Hao Zhang}

\author{Linjie Zhang}

\author{Liantuan Xiao}

\author{Suotang Jia}
\affiliation{State Key Laboratory of Quantum Optics and Quantum Optics Devices, Institute of Laser Spectroscopy, Shanxi University, Taiyuan, Shanxi 030006, China}
\affiliation{Collaborative Innovation Center of Extreme Optics, Shanxi University, Taiyuan, Shanxi 030006, China}

\date{\today}

\begin{abstract}
The laser's frequency noise is crucial to the sensitivity of quantum sensors. Two commonly used methods to suppress the laser's frequency noise are locking the laser to an atomic transition by the lock-in technique or to an ultra-low thermal expansion (ULE) glass cavity by the PDH technique. The former cannot suppress rapidly changing frequency noise and hardly meets the needs; the latter has powerful performance but a heightened cost. The lack of high-performance and low-cost laser noise suppression methods dramatically limits the practical application of quantum sensors. This work demonstrates that, in many quantum sensing applications such as the Rydberg atomic superheterodyne receiver, by cascade locking the laser to both the atomic transition and a low-cost (LC) cavity, the same performance as locking to the ULE cavity can be achieved. This work is significant in promoting the practical application of quantum sensors.
\end{abstract}
\maketitle

Lasers are essential for realizing quantum sensors. The laser's noise characteristics, especially the frequency noise characteristics, will significantly affect the quantum sensor's sensitivity, as proven in many kinds of quantum sensors\cite{degen2017quantum}, including but not limited to atomic clocks \cite{jiang2011making,oelker2019demonstration}, atomic magnetometers\cite{schultze2010noise}, and Rydberg atomic electric field sensors \cite{jing2020atomic,yang2024sensitivity,osterwalder1999using,sedlacek2012microwave,mohapatra2008giant}.

In order to improve the sensitivity of quantum sensors, the frequency noise of the laser is usually suppressed by locking the laser to a stable frequency reference. According to the choice of the reference frequency, these methods can be roughly divided into two categories: one uses the atomic transition, and the other uses optical cavities or optical frequency combs. Since atomic spectrum, such as saturated absorption (SAS) spectrum \cite{preston1996doppler} and electromagnetically induced transparency (EIT) spectrum \cite{mohapatra2007coherent,boller1991observation}, are easy to obtain, the first frequency noise suppression method has the most extensive applications \cite{wallard1972frequency,mccarron2008modulation,fancher2023self}. However, it takes a relatively long time for the light-atomic interaction to establish a steady state, making it difficult for this method to suppress the high-frequency components of frequency noise. The second higher-performance approach to frequency noise suppression is to lock the laser to an ultra-low thermal expansion glass Fabry–Pérot cavity (ULE-FPI) \cite{matei20171,kedar2023frequency}. In order to ensure long-term frequency stability, this powerful cavity operates at the zero expansion temperature of ULE glass and is placed in a vibration-isolation vacuum environment. These designs make the ULE cavity expensive and consume heightened power during operation. The lack of economical and high-performance frequency noise suppression methods dramatically limits the practical applications of quantum sensors because reducing their size, weight, power, and cost (SWaP-C) is also critical in these applications.

This work realizes a practical, high-performance laser system that can meet the requirements for SWaP-C in practical applications. At the same time, its frequency noise at high Fourier frequencies ($f$) is effectively suppressed, thus meeting the needs of quantum sensors operated at high readout frequencies, e.g., atomic superheterodyne receivers (atomic superhet), which has developed rapidly in recent years.

The laser system first locks the laser to a low-cost cavity (LC-FPI) through the Pound-Drever-Hall (PDH) technique \cite{drever1983laser,black2001introduction}. The LC cavity has significant noise at low frequencies due to the influence of temperature changes, air refractive index changes, and vibration. However, these noises decrease rapidly at a rate of $1/f$ as frequency increases; thus, the LC cavity will soon have frequency noise that is lower than the noise of the laser itself as frequency increases, and locking the laser to the LC cavity can effectively suppress the frequency noise of lasers at high Fourier frequencies. At the same time, the sizeable frequency noise at the low Fourier frequencies of the LC cavity will be transmitted to the laser, which may even cause noise more significant than the noise of the laser itself. In order to overcome this problem, the laser locked to the LC cavity is injected into a saturated absorption spectrum module to generate an error signal, and the feedback is injected into the LC cavity's piezoelectric ceramic (PZT) to lock the LC cavity mode to the atomic transition. Thereby, the low-frequency noise of the LC cavity, and thus the laser, is suppressed. Reducing laser noise by cascade locking to an LC cavity and atomic transition has long been achieved but needs a detailed performance characterization \cite{sedlacek2013atom}. This work shows that this approach can achieve the same performance in many applications as locking the laser to a ULE cavity.

\begin{figure*}
\centering
\includegraphics[width=1\linewidth]{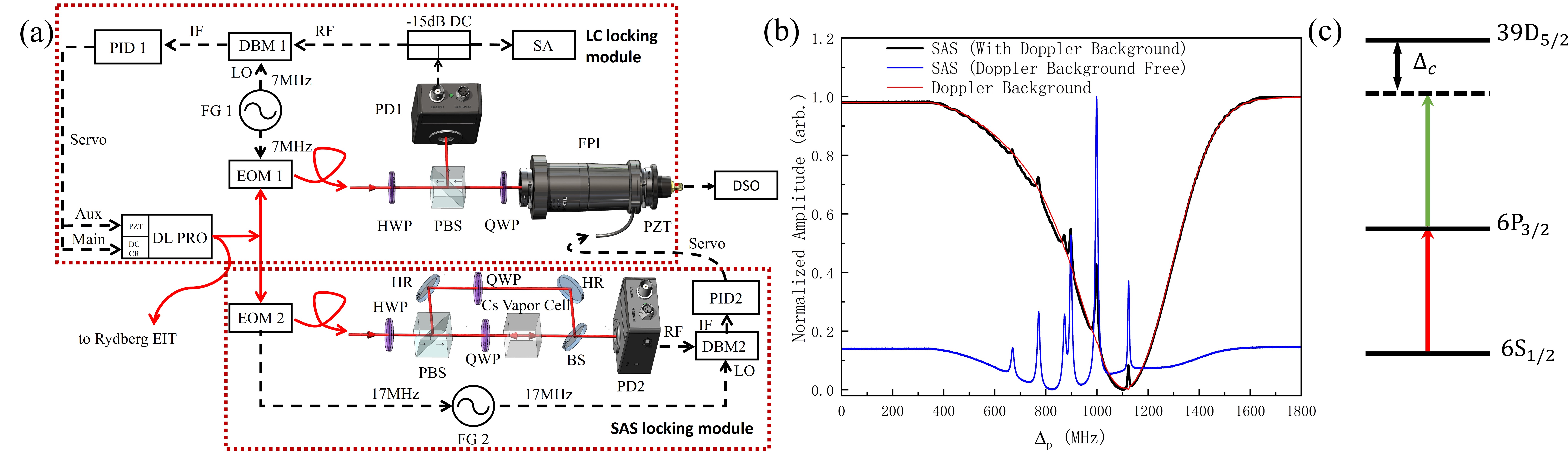}
\caption{\label{fig1} Overall experimental configuration. (a) Experimental setup. Solid lines represent optical fibers, and dashed lines represent circuit connections. We have used the following notations: DL PRO: external-cavity diode laser, DC CR: DC-coupled laser current port, PZT: piezo ceramic, EOM1 \& EOM2: electro-optic modulator, PD1 \& PD2: photodetector, -15 dB DC: -15 dB directional electronics coupler, DBM1 \& DBM2: double balanced mixer, PID1 \& PID2: proportional-integral-derivative controller, SA: spectrum analyzer, FG1 \& FG2: function generator, HWP: half-wave plate, BS: beam splitter (T:R=1:1), HR: high reflection dielectric mirror, PBS: polarizing beam splitter, QWP: quarter-wave plate, FPI: Fabry-Pérot cavity. DSO: digital storage oscilloscope. (b): SAS of Cesium $\rm{D_2}$-transition. The black curve corresponds to the original SAS. The red curve corresponds to the Doppler absorption background. The blue curve corresponds to the Doppler background deduced SAS, and the hyperfine structure is clearly shown. (c): Energy level diagram of Rydberg EIT. The ground state is ${\rm 6S_{1/2},F=4}$, excited state is ${\rm 6P_{3/2},F'=5}$ and Rydberg state is ${\rm 39D_{5/2}}$.
}
\end{figure*}

Figure \ref{fig1} (a) shows the detailed experimental setup. The 852 nm linear polarized laser emitted by the commercial external cavity diode laser (ECDL) is divided into three parts. The first and second parts of the laser, occupying a small part of the total laser power, are injected into the {\bf LC cavity locking module} and the {\bf SAS locking module}, respectively. The third part of the laser can be used to implement the quantum sensor, and in this work, it is used as the probe laser of the Rydberg EIT spectrum to realize an atomic superhet. In the {\bf LC cavity locking module}, we use the PDH technique to lock the laser to a commercial LC cavity (Thorlabs SA30-95) to suppress the frequency noise of the laser at high Fourier frequencies; the LC cavity has a fineness of 1500 and a linewidth of 1 MHz. The optical configuration of the PDH technique is the same as the usual \cite{drever1983laser,black2001introduction}. The PDH sideband is generated by modulating a fiber EOM (EOM1) with a 7 MHz sine wave. The 7 MHz sine wave is provided by one of its ports of a two-channel function generator (FG1), and the other port of FG1 provides the demodulation signal with the same frequency but adjustable phase to the local (LO) port of a double-balanced mixer (DBM1). The modulation signal frequency is set to more than twice the linewidth of the LC cavity for a steep slope of the error signal \cite{black2001introduction}. The laser with PDH sidebands reflected by the LC cavity is converted to an electrical signal by a high bandwidth photodetector (PD1). We use a -15 dB directional electronics coupler to split a small portion (3\%) of the PD1 output signal to monitor the signal with a spectrum analyzer. Most of the PD1 output signal is input to the RF port of the DBM1 to generate an error signal. A PID controller (PID1) processes the error signal into feedback signals; its main output, with a bandwidth of over 10 MHz, is injected into the DC-coupled laser current port to suppress rapidly changing frequency noise; its auxiliary output, with a small bandwidth, is injected into the laser PZT port to compensate for slow but wide-range frequency noise and prevent saturation of the DC current port. In the {\bf SAS locking module}, we lock the frequency of LC cavity mode to Cesium ${\rm 6S_{1/2},F=4~\to~6P_{3/2},F'=5}$ transition via SAS. The laser is divided into probe and pump lasers through the combination of HWP and PBS. The probe laser and pump laser reverse overlap in the Cesium vapor cell, and PD2 detects the probe laser to obtain the SAS (Fig. \ref{fig1} (b)). FG2 applies a sine modulation signal with a frequency of 17 MHz to EOM2 and a sine demodulation signal to the LO port of DBM2 with the same frequency but an adjustable phase. The feedback signal is injected into the PZT of the LC cavity to lock the cavity mode to the same frequency as the atomic transition, thereby suppressing the frequency noise at the low Fourier frequency of the cavity and, thus, the noise of the laser.

\begin{figure*}
\centering
\includegraphics[width=1\linewidth]{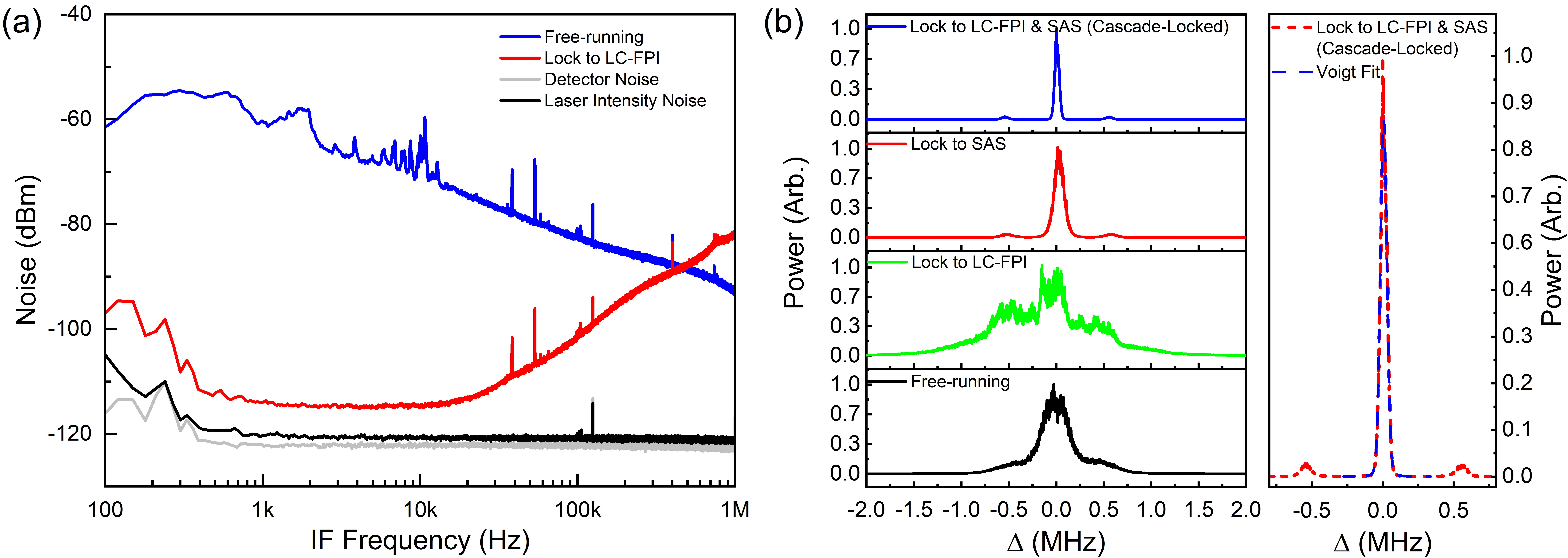}
\caption{\label{fig2} Locking performance analysis. (a) The single-side noise spectrum,  extracted from the small part of the PD1 signal collected by the spectrum analyzer. Gray curve: detector amplifier noise, measured by blocking the optical input of PD1. Black curve: Laser intensity noise, measured by adjusting the laser frequency to be far detuned relative to the cavity mode. Blue curve: relative frequency noise between the free-running laser and the LC cavity, measured by loose locking (limiting the loop bandwidth within 1 kHz by low-pass filter) the laser to the cavity resonance. Red curve: relative frequency noise between the tight-locked laser and the LC cavity. (b) Beat frequency results between laser in this work under different locking configurations (see figure label) and a laser with Hertz-level linewidth (laser locked to a ULE-FPI, the same laser system as in our previous work \cite{jing2020atomic}, and the small peak at $\pm$ 0.55 MHz is caused by the servo pump of its feedback loop).
}
\end{figure*}

Benefiting from the PDH technique, the feedback loop of the LC cavity locking module can achieve a locking bandwidth of over 1 MHz. It can achieve tight locking between laser frequency and LC cavity modes, as shown in Fig. \ref{fig2} (a). The relative frequency noise between the tight-locked laser and LC cavity is suppressed by about 60 dB compared with free-running. However, since the LC cavity itself has considerable low-frequency noise, the linewidth of the tight-locked laser even exceeds the linewidth of the free-running laser, as shown in the black and green curve in the left panel of Fig. \ref{fig2} (b). The low-frequency noise of the tight-locked laser can be suppressed by locking the cavity mode to the atomic transition, and the result is shown in the blue curve in the left panel of Fig. \ref{fig2} (b). The linewidth of the cascade-locked laser is significantly narrowed, which is smaller than that of a free-running laser. In the right panel of Fig. \ref{fig2} (b), the fitting results show that the lineshape of the cascade-locked laser is close to Gaussian with a linewidth of 53 kHz, proving that the linewidth of the cascade-locked laser mainly depends on the unsuppressed low-frequency $1/f$ noise \cite{stephan2005laser}. Fig. \ref{fig2} (b) also shows the beat frequency results of a laser that only locks to atomic transitions (red curve in the left panel). The linewidth is between free-running and cascade-locked lasers, proving that locking the laser to the LC cavity can suppress laser frequency noise.

\begin{figure}
\includegraphics[width=1\linewidth]{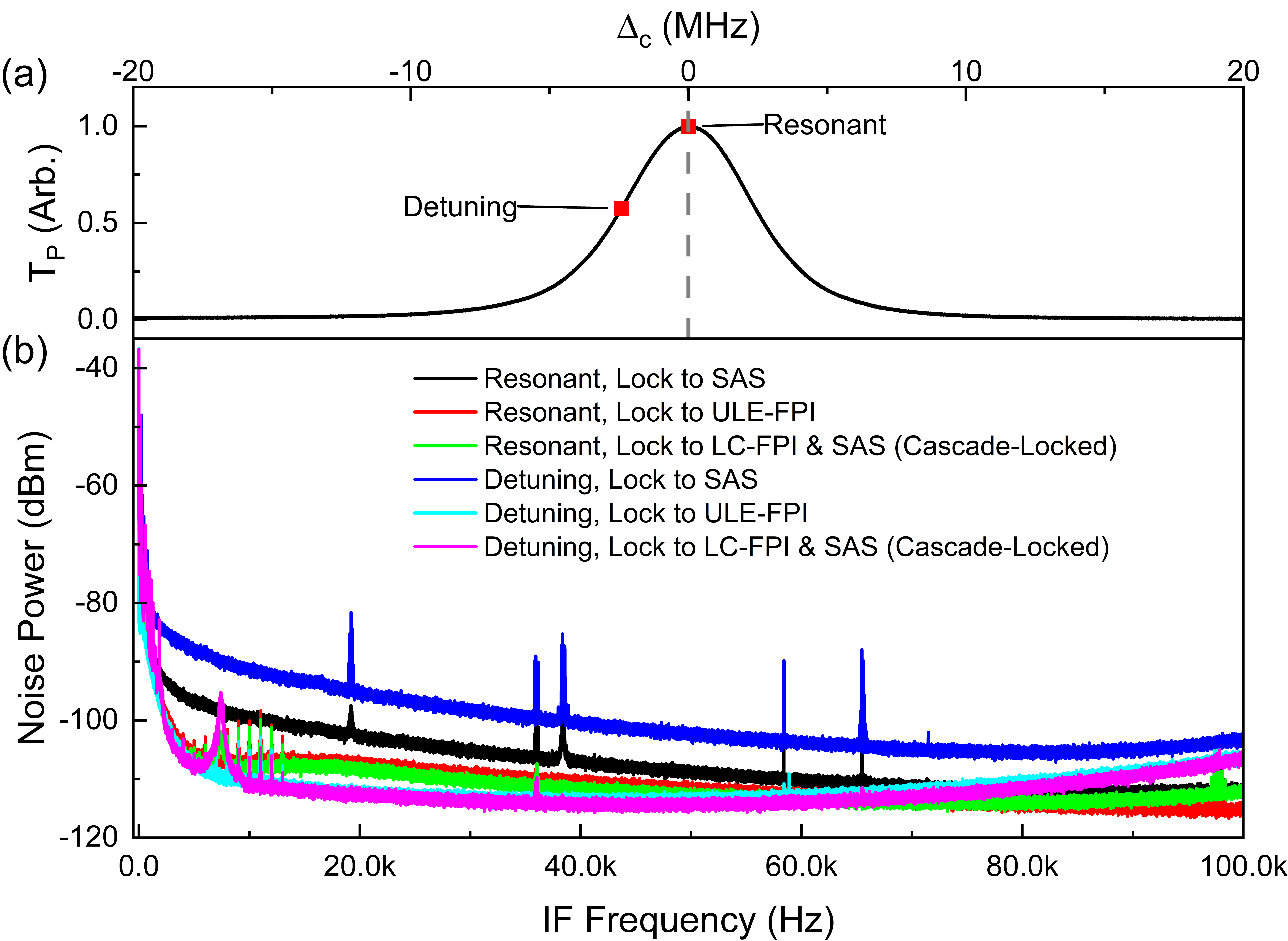}
\caption{\label{fig3} The readout noise spectrum of atomic superhet at different probe laser locking configurations and coupling laser detuning. The probe laser is resonant with the transition frequency of ${\rm 6S_{1/2},F=4~\to~6P_{3/2},F'=5}$, by locking to SAS, ULE-FPI or by cascade-locking to LC-FPI \& SAS. The coupling laser is resonant with or 2.4 MHz detuned from the transition frequency of ${\rm 6P_{3/2},F'=5~\to~39D_{5/2}}$. (a) Example EIT spectrum. The y-axis is probe laser transmission (${\rm T_P}$). The two coupling laser frequency settings are marked as red points. (b) Readout noise spectrum.}
\end{figure}

Linewidth is an integrating result of laser frequency noise and depends primarily on noise at the low Fourier frequencies. From linewidth results, cascade-locked lasers are incomparable with those locked to ULE cavities in low-frequency noise suppression since those locked to ULE cavities can usually achieve sub-Hertz-level linewidths \cite{kedar2023frequency}. However, many quantum sensors do not operate at low readout frequencies and are not concerned with laser frequency noise at low Fourier frequencies. A typical example of such a quantum sensor is an atomic superhet, which performs measurements at readout frequencies over 10 kHz \cite{jing2020atomic}. The measurement results of the atomic superhet are read out through the Rydberg EIT spectrum, so its readout noise can be analyzed through the noise of the Rydberg EIT spectrum \cite{zhang2023quantum}, as shown in Fig. \ref{fig3}. In this work, we use commonly used configurations of the EIT spectrum \cite{jing2020atomic,sedlacek2012microwave}, i.e., the probe laser and coupling laser reverse overlap in a Cesium vapor cell, and their power is actively locked to reduce intensity noise. The probe laser is locked to the transition frequency of ${\rm 6S_{1/2},F=4~\to~6P_{3/2},F'=5}$ by using of different locking configurations. The coupling laser drives the transition of ${\rm 6P_{3/2},F'=5~\to~39D_{5/2}}$ resonantly or with a small detuning, depending on which kind of atomic superhet is implemented. The coupling laser is locked to a high-performance ULE cavity so that its frequency noise does not affect the final readout noise. Atomic superhets can currently be classified into two types, depending on whether the radio wave frequency to be measured is near resonance with the transition frequency between nearby Rydberg levels. In the first type, the radio wave, typically at microwave frequency, interacts resonantly with nearby Rydberg levels, and the coupling laser is kept resonant with the corresponding transition. Since both the probe laser and the coupling laser resonate with the corresponding transitions (see the resonant mark in Fig. \ref{fig3} (a)), the first derivative of the probe laser transmission versus the laser detuning is zero. This type of atomic superheterodyne receiver is insensitive to the laser's frequency noise. The black, red, and green curves in Fig. \ref{fig3} (b) show the readout noise of resonant type atomic superhet with probe laser locked to SAS, ULE cavity, or cascade-locked to LC cavity \& SAS, respectively. The results show that the cascade-locked probe laser performs similarly to the laser locked to the ULE cavity in the readout frequency range of 10 kHz to 100 kHz and up to 10 dB less noise than the laser locked to SAS. In the second type of atomic superhet, the radio wave interacts far-detuning with Rydberg atoms, and the frequency of the coupling laser is kept at the slope of the EIT spectrum to achieve maximum sensitivity \cite{li2023super,liu2022highly}. As in this work, we set the detuning of the coupling laser to 2.4 MHz, situated on the slopes of our EIT spectrum (see the detuning mark in Fig. \ref{fig3} (a)). This type of atomic superhet is much more sensitive to the laser's frequency noise than the first type because the first derivative of the probe laser transmission versus the laser detuning is non-zero. The blue, light gray, and cyan curves in Fig. \ref{fig3} (b) show the readout noise of detuning type atomic superhet with probe laser locked to SAS, ULE cavity, or cascade-locked to LC cavity \& SAS, respectively. The cascade-locked probe laser still performs similarly to the laser locked to the ULE cavity in the readout frequency range of 10 kHz to 100 kHz, and the laser locked to SAS is at least 20 dB worse.

In summary, by locking the laser to a low-cost cavity and cascading locking the cavity to the SAS, we have achieved a practical, high-performance laser system that can be used for quantum sensors. When used in an atomic superhet, the laser can achieve precisely the same performance as locking to an expensive ULE cavity over the readout frequency range of 10 kHz to 100 kHz, far better than just locking it to SAS. In addition to the atomic superhet, this laser system is suitable for any quantum sensors operated at high readout frequencies and insensitive to the laser's frequency noise at low Fourier frequencies. Because frequency noise at low Fourier frequencies is not well suppressed, this laser system cannot compete with lasers locked to the ULE cavity in the range of readout frequency of less than 1 kHz. Therefore, it cannot be applied to low-frequency electric field measurements based on Rydberg atoms \cite{li2023super,jau2020vapor,cox2018quantum}, and this defect still needs to be solved by further research.

\begin{acknowledgments}
This research is funded by the National Key R\&D Program of China (grant no. 2022YFA1404003),  the National Natural Science Foundation of China (grants 12104279, 61827824 and 61975104), Innovation Program for Quantum Science and Technology (Grant No. 2021ZD0302100), Shanxi Provincial Key R\&D Program (202102150101001), the Fund for Shanxi ‘1331 Project’ Key Subjects Construction, Bairen Project of Shanxi Province, China.
\end{acknowledgments}

\section*{AUTHOR DECLARATIONS}
\section*{Conflict of Interest}
The authors have no conflicts to disclose.
\section*{Authors' contributions:}
Authors' contributions: M.J. proposed the project, developed the research and analyzed the experimental data. S.X. and M.J. performed the experiments. M.J., L.Z. and H.Z. contributed to the experimental setup. S.X. and M.J. wrote the manuscript. All authors contributed to discussions of the results and the manuscript and provided revisions of manuscript.
\section*{DATA AVAILABILITY}
The data that support the findings of this study are available from the corresponding author upon reasonable request.

\section*{REFERENCES}
\bibliography{reference.bib}

\end{document}